\documentclass[journal]{IEEEtran}

\usepackage{enumerate}
\usepackage{amsmath,amssymb,amsthm,amsfonts,mathrsfs}
\usepackage{mathtools}
\usepackage{bbold} 
\usepackage{optidef}
\usepackage{cite}
\usepackage{pgfplots}
\pgfplotsset{compat=newest}
\usepackage[utf8]{inputenc}

\usepackage{hyperref}
\usepackage{stmaryrd}
\usepackage{bbm}
\usepackage{cuted}
\usepackage{quantikz}

\usepackage{graphicx,color,cases} 
\usepackage{xfrac}

\definecolor{DarkGreen}{rgb}{0.1,0.5,0.1}
\definecolor{DarkRed}{rgb}{0.5,0.1,0.1}
\definecolor{DarkBlue}{rgb}{0.1,0.1,0.5}

\allowdisplaybreaks

\newtheorem{theorem}{Theorem} 
\newtheorem{lemma}[theorem]{Lemma}

\theoremstyle{definition}

\newtheorem{remark}{Remark}
\numberwithin{equation}{section}
 
\def\>{\rangle} 
\def\<{\langle}

\newif\ifcomment
\commenttrue

\begin{document}

\title{Explaining Robust Quantum Metrology by Counting Codewords}
\author{Oskar~Novak,~\IEEEmembership{Graduate Student Member,~IEEE,}
        and Narayanan~Rengaswamy,~\IEEEmembership{Senior Member,~IEEE}        
\thanks{Oskar Novak and Narayanan Rengaswamy are with the Department of Electrical and Computer Engineering, University of Arizona, Tucson, AZ, 85721 USA. E-mail: \{ onovak , narayananr \}@arizona.edu}}


\maketitle

\begin{abstract}
Quantum sensing holds great promise for high-precision magnetic field measurements. 
However, its performance is significantly limited by noise.
The investigation of active quantum error correction to address this noise led to the Hamiltonian-Not-in-Lindblad-Span (HNLS) condition.
This states that Heisenberg scaling is achievable if and only if the signal Hamiltonian is orthogonal to the span of the Lindblad operators describing the noise.
In this work, we consider a robust quantum metrology setting where the probe state is inspired from CSS codes for noise resilience but there is no active error correction performed.
After the state picks up the signal,  we measure the code’s $\hat{X}$ stabilizers to infer the magnetic field parameter, $\theta$.
Given $N$ copies of the probe state, we derive the probability that all stabilizer measurements return $+1$, which depends on $\theta$. 
The uncertainty in $\theta$ (estimated from these measurements) is bounded by a new quantity, the \textit{Robustness Bound}, which ties the Quantum Fisher Information of the measurement to the number of weight-$2$ codewords of the dual code.
Through this novel lens of coding theory, we show that for nontrivial CSS code states the HNLS condition still governs the Heisenberg scaling in our robust metrology setting.
Our finding suggests fundamental limitations in the use of linear quantum codes for dephased magnetic field sensing applications both in the near-term robust sensing regime and in the long-term fault tolerant era.
We also extend our results to general scenarios beyond dephased magnetic field sensing.
\end{abstract}


\section{Introduction}
\label{sec:intro}

 \IEEEPARstart{Q}{uantum} sensors have long been known to outperform classical ones by reaching a tighter uncertainty bound called the Heisenberg Limit. 
 However, the Heisenberg Limit remains elusive in the presence of noise. 
 Many approaches to circumvent this problem utilize active quantum error correction to suppress the channel's noise~\cite{Kapourniotis-pra19,Reiter-natcomm17,Rojkov-prl22,Kessler-prl14,Dur-prl14,Layden-prl19,Unden-prl16}. 
 There have also been some error detection approaches~\cite{Matsuzaki-pra17}. 
 Non-Markovian resources and non-linear quantum systems have also been widely considered for metrology applications~\cite{Yang-commphy24,Beau-prl17}. 
 However, the active quantum error correction approach has downfalls. 
 There is a strong no-go theorem for infinite rounds of active error correction that prevents surpassing the Standard Quantum Limit if the Hamiltonian of the system is in the span of the Lindblad operators~\cite{Zhou-natcomm18}.
 In other words, the Heisenberg Limit is achievable if and only if the Hamiltonian is Not in the Lindblad Span, known as the HNLS condition. 
 If HNLS is violated, then the noise becomes indistinguishable from the Hamiltonian, causing active error correction to suppress part of the signal inadvertently. 
 
 Heisenberg Scaling is achievable by using Dicke States~\cite{Saleem-pra24}, which can be considered as superpositions of the codewords of some non-linear error correcting code. 
 In practice, constructing Dicke states remains challenging, and with the fault-tolerant era of quantum computing on the horizon, leveraging logical states of linear (stabilizer) codes presents a more feasible long-term approach. 
 This approach can be further simplified by employing \emph{robust} metrology in the near term, where we encode the probe state into a logical state of a stabilizer code \emph{but do not perform active error correction}.
 Such a strategy can easily adapt to the fault tolerant era once the technology is mature.
 This robust metrology approach using stabilizer codes was shown to achieve the Heisenberg Limit under erasure noise~\cite{Ouyang-pra23}, which motivates us to extend beyond erasures to the general setting comprised of $\hat{X}$- and $\hat{Z}$-noise. 
 
 We consider a robust quantum metrology protocol that prepares a probe state as the superposition of codewords of a classical linear code $C$. 
The state can be reinterpreted as the logical $\ket{00\cdots 0}$ state of a CSS code whose $\hat{X}$-stabilizers are described by $C$.
After encoding the parameter into the state through evolution under the sensing Hamiltonian for time $dt$, the protocol measures these $\hat{X}$-stabilizers to determine $\theta$. 
The evolution suffers from dephasing noise, which is relevant for robust quantum metrology~\cite{Ouyang-it22}. 
We note that in Ref.~\cite{Ouyang-pra14}, the setup was limited to permutation-invariant codes, which are non-linear, where the probe state undergoes a \emph{fixed} amount of dephasing errors. 
 Furthermore, the only requirement on the Quantum Fisher Information (QFI) was that it remained positive after undergoing the noise channel. 

 Here, our primary contribution is to employ a novel coding-theoretic approach to investigate the performance of CSS code states in robust quantum metrology. Using this approach we show that the QFI of the measurement is governed by a new quantity called the \textit{Robustness Bound}. This bound ties the scaling behavior of the QFI induced by the probe state to the weight enumerator of the dual of the code $C$. 
 This bound demonstrates that Heisenberg Scaling is only achievable for non-trivial CSS codes when the Hamiltonian is orthogonal to the span of the Lindblad operators, a result that extends the HNLS condition to the robust metrology case. 
 
 In Sec.~\ref{sec:toyexample}, we first consider a known set of $\hat{Z}$-errors, similar to Ref.~\cite{Ouyang-pra14} except that we use CSS code states rather than non-linear code states.
 We use this as a toy example to build intuition for our usage of the weight enumerator of the dual code of $C$ to derive the QFI of the probe state.
 Then we consider a full-rank dephasing channel in Sec.~\ref{sec:protocol} and derive the main results in Sec.~\ref{sec:results}.
 Our results demonstrate an intriguing connection between coding theory and quantum metrology not yet appreciated in the literature. We show that by simply considering the structure of linear error-correcting codes, we can derive fundamental physical results in an elegant and intuitive fashion. 
 Finally, we complete the picture by considering general Pauli noise that is a mixture of $\hat{Z}$ and $\hat{X}$ errors, where we show that Heisenberg Scaling is achievable only under purely $\hat{X}$-errors.

 The paper is organized as follows: Section~\ref{sec:background} introduces the necessary background for our paper, Section~\ref{sec:toyexample} motivates the usage of the weight enumerator of the dual code for deriving sensing bounds via a toy example, Section~\ref{sec:protocol} details our sensing protocol, Section~\ref{sec:results} introduces our main results, and Section~\ref{sec:outlook} concludes by discussing the outlook given our results. 
 The appendices provide the proofs of the main results.

\section{Background}
\label{sec:background}

\subsection{Single Parameter Quantum Metrology}
\label{sec:spqm}

We wish to sense a classical parameter $\theta$ related to the strength of a weak magnetic field. In this case, the quantum probe state $\rho$ evolves with respect to the Hamiltonian:  
\begin{equation}
    \label{hamiltonian}
    \hat{H}=\sum_{j=1}^N \hat{Z}_{j}.
\end{equation}

We wish to sense $\theta$ with as little uncertainty as possible, meaning we wish to minimize the standard deviation: 
\begin{equation}
    \label{CRB}
    \delta \theta \geq \frac{1}{\sqrt{\mathcal{Q}(\rho,\hat{O}_{\theta})}},
\end{equation}
where~\eqref{CRB} is the Quantum Cramer-Rao bound~\cite{Braunstein-prl94} and $\mathcal{Q}(\rho,\hat{O}_{\theta})$ is the Quantum Fisher Information (QFI), defined as follows with respect to some measurement operator $\hat{O}_{\theta}$~\cite{Helstrom-jsp69}:
\begin{equation}
    \label{qfi}
    \mathcal{Q}(\rho,\hat{O}_{\theta}) \coloneqq 2\sum_{k,l} \frac{(\lambda_k-\lambda_l)^{2}}{\lambda_{k}+\lambda_{l}}|\bra{k}\hat{O}_{\theta}\ket{l}|^2.
\end{equation}
Here, $\{\lambda_{i}\}$ and $\{\ket{i}\}$ are the set of eigenvalues and eigenvectors of $\rho$, respectively. An optimal choice of measurement operator for determining~\eqref{qfi} is the Symmetric Logarithmic Derivative (SLD), $\hat{L}_{\theta}$, defined as~\cite{Braunstein-prl94}:
\begin{equation}
\label{sld}
\frac{\partial \rho}{\partial \theta} \coloneqq \frac{1}{2} \{\rho,\hat{L}_{\theta}\}.
\end{equation}
There are two limits of \eqref{CRB} that are relevant to our goal: the Heisenberg Limit (HL), given by $\delta \theta \geq \frac{1}{N}$, which is the minimal uncertainty allowed by Quantum Mechanics, and the Standard Quantum Limit, (SQL), given by $\delta \theta \geq \frac{1}{\sqrt{N}}$, where $N$ is the number of qubits of the probe state. When the system is noiseless, it can be shown that the $N$-qubit GHZ state gives the optimal probe state for achieving the HL~\cite{Eldredge-pra18}:
\begin{equation}
    \label{ghz}
    \ket{\textrm{GHZ}} \coloneqq \frac{1}{\sqrt{2}}\ket{0}^{\otimes N}+\frac{1}{\sqrt{2}}\ket{1}^{\otimes N}.
\end{equation}
However, this is no longer true when the system is noisy. 
Thus, to maximize the QFI, we must first quantify the noisy evolution of the probe state.

\subsection{Lindbladian Master Equation}
\label{sec:linblad}

The probe state undergoes noisy evolution, which is non-unitary. In general, the noisy evolution of a density operator is given by some Kraus Operators $\{ M_{k} \}$~\cite{Kraus-83} such that:
\begin{align}
    \label{kraus}
    &\nonumber \rho \rightarrow \sum_{k} M_{k}\rho M_{k}^{\dagger},
    \\ &\sum_{k} M_{k}^{\dagger} M_{k} = I.
\end{align}
Given a unitary operator $U(t,\theta)=e^{\frac{-i}{\hbar} \hat{H}(\theta) t}$, we have
\begin{equation}
\label{unitary op evol of rho}
    \rho(t)=U(t,\theta) \, \rho\, U^{\dagger}(t,\theta).
\end{equation}
This is the solution of the differential equation
\begin{equation}
\label{LVN}
    \frac{\partial \rho}{\partial t}=-\frac{i}{\hbar} [\hat{H}(\theta),\rho],
\end{equation}
known as the \textit{Liouville-Von Neumann} equation. 
Inserting \eqref{kraus} into \eqref{LVN}, for a Markovian channel, we derive the Lindblad Master-Equation~\cite{Breuer-02}:
\begin{align}
    \label{Linbladian}
    & \nonumber\frac{\partial \rho}{\partial t}=-\frac{i}{\hbar}  [\hat{H},\rho]\theta+\sum_{k}\left(L_{k}\rho L_{k}^{ \dagger}-\frac{1}{2}\{L_{k}^{\dagger}L_{k},\rho\}\right).
\end{align}
Here, we can define our Lindblad (jump) operators $L_{k}=\frac{1}{\sqrt{dt}}M_{k}, k\geq 1$, and we let $\hat{H}(\theta)=\theta\hat{H}$.

\subsection{Code Weight Enumerator}
\label{sec:codeweight}

To maximize the QFI of measurements in the presence of noise, we define the probe state to be the equally weighted superposition of codewords of a classical code $C$.
The probe state can be reinterpreted as the logical $\ket{00\cdots 0}$ state of a CSS code with $\hat{X}$-stabilizers defined by $C$.

Since we use superpositions of codewords of some code $C$ as the probe state, it will be necessary to enumerate the number of codewords of Hamming weight $k$, for all $k \in \{ 0,1,\ldots,n \}$. 
The weight corresponds to the number of Pauli $\hat{X}$ operators in an $\hat{X}$-stabilizer derived from a codeword of $C$. 
The weight enumerator is a polynomial whose coefficients are the number of codewords of each weight $k$ in the code $C$:
\begin{equation}
\label{enumerator}
W_{C}(z) \coloneqq \sum_{k=0}^{N} W_{C,k} z^{k},
\end{equation}
where $z$ is some formal variable~\cite{Hill-86}. 
As we will show later, the weight enumerator of the dual code of $C$, denoted $C^{\perp}$, will determine the QFI. 
We can find the dual code's weight enumerator $W_{C^{\perp}}(z)$ by using the MacWilliams identity~\cite{Macwilliams-77}:
\begin{equation}
     \label{eq:mcwilliams}
     W_{C^{\perp}}(z) = \frac{1}{|C|}(1+z)^{N}W_{C}\left(\frac{1-z}{1+z}\right).
 \end{equation}
 To minimize the uncertainty of the measurement, we must choose a probe state $\rho$ that maximizes QFI, which will depend on \eqref{eq:mcwilliams}. 
 We will show that it suffices to choose a code $C$ whose dual code has many codewords of a certain weight $k$, i.e., that such a code maximizes the QFI for the channel. 
 To demonstrate this, next we consider a toy example that provides intuition for how the QFI is related to \eqref{eq:mcwilliams}.

 \section{Finding optimal probe state under a fixed number of $\hat{Z}$-errors }
 \label{sec:toyexample}

Let $\ket{\psi}=\frac{1}{\sqrt{|C|}}\sum_{\boldsymbol{x}\in C}\ket{\boldsymbol{x}}$ be an evenly weighted superposition of all the classical codewords in a binary linear code $C$, where $\boldsymbol{x} \in \{0,1\}^{N}$. We can write the pure state in density operator form as: 
\begin{equation}
\label{eq:rho}
    \rho=\frac{1}{|C|}\sum_{\boldsymbol{x},\boldsymbol{y} \in C}\ket{\boldsymbol{x}}\bra{\boldsymbol{y}}.
\end{equation}
Consider the channel that applies a $Z$ error to exactly $w \leq N$ qubits, with all $w$-qubit errors being equally likely. The resulting probe state can be written in Kraus representation as:
\begin{equation}
\varepsilon(\rho)=\binom{N}{w}^{-1}\sum_{\boldsymbol{i}\in \mathcal{I}} E(\boldsymbol{0},\boldsymbol{i}) \rho E(\boldsymbol{0},\boldsymbol{i}).
\end{equation}
Here, $E(\boldsymbol{0},\boldsymbol{i})$ is a special case of the Pauli operator:
\begin{equation}
\label{shorthandE}
E(\boldsymbol{a},\boldsymbol{b}) = \imath^{a_{1}b_{1}}\hat{X}^{a_{1}}\hat{Z}^{b_{1}}\otimes...\otimes \imath^{a_{N}b_{N}}\hat{X}^{a_{N}}\hat{Z}^{b_{N}}, 
\end{equation}
where $\boldsymbol{a}=[a_1,a_2,\ldots,a_N], \boldsymbol{b} = [b_1,b_2,\ldots,b_N] \in \{0,1\}^{N}$, $\imath = \sqrt{-1}$, $|\boldsymbol{i}|=w$, and $\mathcal{I}$ is set of all length $N$ binary vectors with weight $w$. 
Let $\langle \cdot , \cdot \rangle$ denote the binary inner product.
Since $Z^u \ket{v} = E(0,u) \ket{v} = (-1)^{uv} \ket{v}$ for $u,v \in \{0,1\}$, it is easy to see that $E(\boldsymbol{0},\boldsymbol{i}) \ket{\boldsymbol{x}} = (-1)^{\langle\boldsymbol{x},\boldsymbol{i}\rangle}  \ket{\boldsymbol{x}}$.
Substituting $\rho$ from~\eqref{eq:rho} yields:
\begin{equation}
\label{purechannel}
    \varepsilon(\rho)=\frac{\binom{N}{w}^{-1}}{|C|}\sum_{\boldsymbol{i}\in \mathcal{I},\boldsymbol{x},\boldsymbol{y} \in C} (-1)^{\langle\boldsymbol{x}\oplus \boldsymbol{y},\boldsymbol{i}\rangle}  \ket{\boldsymbol{x}}\bra{\boldsymbol{y}},
\end{equation}
where $\oplus$ is addition over $\mathbb{Z}_{2}$. 
We can use the following QFI upper bound to look at the probe's behavior, which is tight for a pure state~\cite{Sidhu-avsqs20,Ouyang-pra23}:
\begin{equation}
    \label{qfi bound}
    \mathcal{Q}\left(\rho_\theta,\frac{\partial \rho_\theta}{\partial \theta}\right) \leq 4\mathrm{Tr}\left(\rho \hat{H}^{2}\right) - 4\left(\mathrm{Tr}\left(\rho \hat{H}\right)\right)^{2},
\end{equation}
where the Hamiltonian is given by \eqref{hamiltonian}, and  $\rho_\theta=U(\theta)\rho U(\theta)^{\dagger}$ with $U(\theta)=e^{-\frac{\imath}{\hbar}\theta \hat{H}}$. 
Substituting \eqref{purechannel} for $\rho$ into \eqref{qfi bound}, the first term of the RHS is calculated as: 
\begin{align}
& 4\mathrm{Tr} \left(\varepsilon(\rho) \hat{H}^{2}\right) \nonumber \\ 
&= \frac{4 \binom{N}{w}^{-1}}{|C|} \mathrm{Tr} \left( \sum_{\substack{\boldsymbol{i}\in \mathcal{I},\\ \boldsymbol{x},\boldsymbol{y} \in C}} (-1)^{\langle\boldsymbol{x}\oplus \boldsymbol{y},\boldsymbol{i}\rangle}  \ket{\boldsymbol{x}}\bra{\boldsymbol{y}} \sum_{k=1}^N \hat{Z}_{k} \sum_{l=1}^N \hat{Z}_{l}\right) \\
&= \frac{4 \binom{N}{w}^{-1}}{|C|} \mathrm{Tr} \left( \sum_{\substack{\boldsymbol{k}, \boldsymbol{l},\\ \boldsymbol{i}\in \mathcal{I},\\ \boldsymbol{x},\boldsymbol{y} \in C}} (-1)^{\langle\boldsymbol{x}\oplus \boldsymbol{y},\boldsymbol{i}\rangle}  \ket{\boldsymbol{x}}\bra{\boldsymbol{y}} E(\boldsymbol{0},\boldsymbol{k} \oplus \boldsymbol{l}) \right)  \\
&= \frac{4 \binom{N}{w}^{-1}}{|C|}\sum_{\substack{\boldsymbol{k}, \boldsymbol{l},\\ \boldsymbol{i} \in \mathcal{I},\\ \boldsymbol{x},\boldsymbol{y} \in C}} (-1)^{\langle\boldsymbol{x}\oplus \boldsymbol{y},\boldsymbol{i}\rangle} \bra{\boldsymbol{y}} E(\boldsymbol{0},\boldsymbol{k} \oplus \boldsymbol{l}) \ket{\boldsymbol{x}},
\end{align}
 where $\boldsymbol{k}, \boldsymbol{l}$ are binary vectors with a single $1$ in index $k$ and $l$, respectively.
 Note that $E(\boldsymbol{0},\boldsymbol{k} \oplus \boldsymbol{l}) = \hat{Z}_{k} \hat{Z}_{l}$.
Due to the orthonormality of $\ket{\boldsymbol{x}}, \ket{\boldsymbol{y}}$ and the relation $E(\boldsymbol{0},\boldsymbol{k} \oplus \boldsymbol{l}) \ket{\boldsymbol{x}} = (-1)^{\langle \boldsymbol{x},\boldsymbol{k}\oplus\boldsymbol{l}\rangle} \ket{\boldsymbol{x}}$, we have the summand $\bra{\boldsymbol{y}} E(\boldsymbol{0},\boldsymbol{k} \oplus \boldsymbol{l}) \ket{\boldsymbol{x}} = (-1)^{\langle \boldsymbol{x},\boldsymbol{k}\oplus\boldsymbol{l}\rangle} \delta_{ \boldsymbol{x}, \boldsymbol{y} }$.
Enforcing $\boldsymbol{x} = \boldsymbol{y}$, which trivializes the factor $(-1)^{\langle\boldsymbol{x}\oplus \boldsymbol{y},\boldsymbol{i}\rangle}$ and turns the sum over $\mathcal{I}$ into the factor $|\mathcal{I}| = \binom{N}{w}$, the expression reduces to:
 \begin{align}
  4\mathrm{Tr}\left(\varepsilon(\rho) \hat{H}^{2}\right) &= \frac{4}{|C|}\sum_{\boldsymbol{k},\boldsymbol{l},\boldsymbol{x} \in C} (-1)^{\langle \boldsymbol{x},\boldsymbol{k}\oplus\boldsymbol{l}\rangle}  \\
  &= 
 \begin{cases} 
\frac{4}{|C|} \sum_{\boldsymbol{k},\boldsymbol{l}} |C| & \text{if}\ \boldsymbol{k}+\boldsymbol{l} \in C^{\perp}, \\
         0 & \text{if}\ \boldsymbol{k}+\boldsymbol{l} \not\in C^{\perp}
\end{cases}  \\ 
  &= 4 \sum_{\substack{\boldsymbol{k},\boldsymbol{l}: \boldsymbol{k}+\boldsymbol{l} \in C^{\perp}}} (1) \\
  &= 4 (2 W_{C^\perp,2} + N).
\end{align}
Thus, this term is maximized by having many weight-$2$ dual codewords.  
 Similarly, one finds that:
 \begin{align}
     4\left(\mathrm{Tr}\left(\varepsilon(\rho) \hat{H}\right)\right)^{2} &= \left(\frac{4}{|C|} \sum_{\boldsymbol{x} \in C, \boldsymbol{k}} (-1)^{\langle\boldsymbol{x},\boldsymbol{k}\rangle}\right)^{2} \nonumber \\
     &= \frac{16}{|C|^2} \left( \sum_{\boldsymbol{k} \in C^{\perp}} |C| \right)^2,
 \end{align} 
 where we now wish to minimize the number of weight-$1$ dual codewords. 
 Observe that such dual codewords correspond to bits that are always zero for all $\boldsymbol{x} \in C$.
 For any non-trivial code $C$, no such bits will exist, so the above term vanishes.
 Hence, the QFI upper bound can be maximized via expanding \eqref{eq:mcwilliams} for the $k=2$ term. 
 We see that to leading order, we can achieve $W_{C^{\perp},2}= \binom{N}{2}$, which would give us Heisenberg Scaling in the pure case. Thus, for our probe state constructed as above from a non-trivial code $C$, the QFI is given by:
 \begin{align}
     \label{QFIpurs}
     \mathcal{Q}\left(\rho_{\theta},\frac{\partial \rho_{\theta}}{\partial \theta}\right) \leq 4\left(2W_{C^{\perp},2}+N\right)=\mathcal{Q}_{\rm pure}.
 \end{align}
 
 \begin{remark}
 The bound is saturated when $\rho_{\theta}$ is pure, i.e., when $\mathcal{I} = \emptyset$.
 This shows why the GHZ state is optimal for pure unitary channels (noiseless case), since $C$ is the repetition code in that case and, therefore, $C^{\perp}$ has all even-weight vectors. Furthermore, this shows that it suffices to consider only one term of the dual weight enumerator of the code $C$ for calculating the QFI bound of the probe state, which greatly reduces the difficulty of deriving an upper bound such as this one. 
  Even if we assumed $w=N$, since each error is equally likely, this calculation still holds. However, if the probability of errors has dependence on $\theta$, then this no longer holds.   
 \end{remark}

\section{Robust Quantum Sensing Protocol}
\label{sec:protocol}
\begin{figure}[t]
\centering
\includegraphics[page=1]{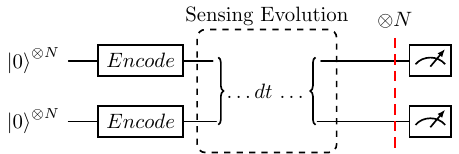}
\caption{The circuit describing our robust sensing protocol. First, we prepare $N$ copies of the encoded $N$-qubit probe state $\rho$. Then, each copy evolves due to the sensing Hamiltonian for time $dt$. Finally, all the $\hat{X}$ stabilizers are measured. This is repeated for several increasing time steps.}
\label{fig:protocol}
\end{figure}

To extend our technique of using the dual weight enumerator to a more general channel, we introduce the following sensing protocol. Typically, the optimal measurement for finding $\theta$ in quantum metrology is given by Eqn.~\eqref{sld}. However, since the eigenvectors of Eqn.~\eqref{sld} typically depend on $\theta$, it is challenging to turn this measurement into an explicit measurement that can be implemented in a lab. 
We find that instead, it is sufficient to measure the syndrome of all the $\hat{X}$-stabilizers of our code after the probe state has evolved for some time $t+dt$. 
We can then calculate the probability of measuring all $+1$ for our syndrome, $p\left(\Pi_{\boldsymbol{1}}|\rho(t+dt)\right)$, by using $N$ copies of the $N$-qubit probe state, where $\Pi$ is the projector constructed from all $\hat{X}$ stabilizers of our code and the subscript corresponds to the syndrome.
Repeating this for a sufficient number of longer time steps, we show in Appendix \ref{sec:AppenA} that our probability will be approximately given by: 
\begin{align}
\label{probabilitymeasuresec4}
p (\Pi_{\boldsymbol{1}} & | \rho(t + dt) ) \approx \nonumber \\
&\frac{1}{2}\left(e^{-\gamma(\theta)(t+dt)}\cos(\sqrt{\mathcal{Q}_{\rm pure}}\theta (t+dt)) +1\right),
\end{align}
which holds to second-order for general codes, but is exact for a GHZ state.
Here, for the case of $N$-qubit dephasing noise acting independently on each of the $N$ qubits of the probe state with probability $p(\theta) \approx p \theta$, we have:
\begin{align}
\label{dampening}
\gamma(\theta) &= 2\left[ 1- \left( (1-p\theta)^N + \tilde{W}_{C^{\perp}} \right) \right] \\
  &= 2 - 2 (1-p\theta)^N W_{C^{\perp}}\left( \frac{p\theta}{1 - p\theta} \right),
\end{align}
where $\tilde{W}_{C^{\perp}}$ is the \emph{robustness} of the code $C$, given by:
\begin{align}
    \label{robustness}
    \tilde{W}_{C^{\perp}} & \coloneqq \sum_{k>0}^{N} (1-p\theta)^{N-k}(p\theta)^{k} W_{C^{\perp},k} \\
    &= (1-p\theta)^N \left[ W_{C^{\perp}}\left( \frac{p\theta}{1 - p\theta} \right) - 1 \right].
\end{align}
Thus, using the setup in Fig.~\ref{fig:protocol} with sufficient time steps, we can recover $\theta$ from the Fourier transform of our time-dependent signal, $p\left(\Pi_{\boldsymbol{1}}|\rho(t+dt)\right)$. 
Recall that for $t \geq 0$, the Fourier transform of $e^{-at} \cos(\omega_0 t)$ is $\frac{a + \omega}{(a + \omega)^2 + \omega_0^2}$.
Therefore, we can sense $\theta$ without relying on rapid and precise control operations at the expense of more significant spatial overhead.

\subsection{Practical Protocol Example}
We provide an example to illustrate our protocol: we may assume that we have an array of $N^{2}$ qubits, such as nitrogen-vacancy (NV) centers, spatially localized so that the difference in the strength of the magnetic field that each qubit experiences is negligible. We then partition the array into $N$ copies of an $N$-qubit GHZ state. Once the state is prepared, the array evolves with respect to the noisy magnetic field system, where the evolution for each GHZ state is governed by ~\eqref{eq:nqcasemain}. After evolving each state for some fixed time $t$, we measure the expectation value of the  $\hat{X}$ stabilizer generators of the probe state $G_{\hat{X}}$, and take the product  of the +1 eigenspace projector of each generator. This is equal to the probability of all stabilizers returning a +1 syndrome at time $t$. In the case of our protocol and probe state, this can be calculated by measuring the $E(\boldsymbol{1},\boldsymbol{0}) = \hat{X}^{\otimes N}$ stabilizer of each GHZ state, where $\boldsymbol{1}=[1,1,\cdots,1]$, and then counting the fraction of states returning +1 for the measurement syndrome. We then repeat this protocol for multiple time steps, until we have sampled enough time steps to clearly extract how the probability distribution $p\left(\Pi_{\boldsymbol{1}}|\rho(t+dt)\right)$ varies with $\theta$. Besides no active error correction, the main difference between our protocol and the typical approach is that, the probability distribution is not calculated by probing with the same probe state multiple times, resetting the probe state, and then evolving the probe again for a longer time $t$. Instead, it is calculated by evolving multiple probe states \emph{simultaneously} for varying times $t$, reducing the amount of time necessary to obtain samples through (spatial) parallelism. 

  \section{Main Result}
  \label{sec:results}

 For the $N$-qubit dephasing channel in Section~\ref{sec:protocol}, we find a new bound called the \textit{Robustness bound}, which demonstrates that the weight enumerator of the dual code encapsulated in \eqref{robustness} determines the scaling behavior of the precision of the probe state measurement. Given this bound, we derive the following theorem:
\begin{theorem} 
\label{Theorem1}
Given an $N$-qubit dephasing channel affecting an all-$\hat{Z}$ Hamiltonian, no non-trivial linear code, $C$, can achieve Heisenberg Scaling for robust field sensing. 
\end{theorem}
The dephasing channel considered here is of the form:
\begin{equation}
\begin{aligned}
 \label{eq:nqcasemain}
 &\frac{\partial \rho}{\partial t}=-\frac{i}{\hbar}[\hat{H},\rho]\theta - \rho+\\&\sum_{k=0, \boldsymbol{j}\in \mathcal{I}}^N (1-p(\theta))^{N-k}(p(\theta))^{k} E(\boldsymbol{0},\boldsymbol{j}) \rho E(\boldsymbol{0},\boldsymbol{j}) ,
 \end{aligned}
 \end{equation}
where $\boldsymbol{j}\in \{0,1\}^N, |\boldsymbol{j}|=k$, $\mathcal{I}$ 
 is the set of all length $N$ binary vectors with weight $k$, and the $-\rho$ term comes from the anti-commutator term from the Lindbladian. We assume that the noise rate is $p(\theta) \approx p \theta$. Here, we find that for the QFI of our measurement to be maximized, we need
\begin{equation}
    \label{new limit prev}
    \tilde{W}_{C^{\perp}} \leq 1-(1-p\theta)^{N} = \sum_{k=1}^N \binom{N}{k} (1-p\theta)^{N-k}(p\theta)^{k}.
\end{equation}
Specifically, saturating this \emph{Robustness Bound} achieves the Heisenberg Scaling. 
But due to the form of $\eqref{new limit prev}$, only the trivial code with dual code consisting of all weight-$1$ vectors saturates the bound, implying that the probe state must be trivial (corresponding to the trivial code $C = \{ 00\cdots 0 \}$).
The proof is given in  Appendix~\ref{sec:AppenA}. 

We now turn to the case where the noise channel is orthogonal to the Hamiltonian. 
For this case, we give the following positive result:
\begin{theorem} 
\label{Theorem2}
Given an $N$-qubit bit-flip channel acting on an all-$\hat{Z}$ Hamiltonian, a non-trivial linear code with maximal number of weight-$2$ dual codewords can asymptotically achieve Heisenberg Scaling for robust field sensing. 
\end{theorem}
In this case, the probe state is the GHZ state and $\gamma(\theta)=0$ for the code. The proof is given in  Appendix~\ref{sec:AppenB}. 

Finally, for the general case, where we have a full-rank channel that mixes the dephasing and bit-flip channels by some polar angle $\phi$, we show that:
\begin{theorem}[Impossibility of Heisenberg-Scaling for Noisy Linear Probes]
\label{Theorem3}
Given an $N$-qubit channel acting on an all-$\hat{Z}$ Hamiltonian such that the channel is comprised of tensor products of the operator $\hat{U}(\phi)=\cos\left(\frac{\phi}{2}\right)\hat{Z}+\sin\left(\frac{\phi}{2}\right)\hat{X}$, a non-trivial linear code can achieve Heisenberg Scaling for robust field sensing if and only if $\phi=\pi$. 
\end{theorem}
\begin{IEEEproof}
It is sufficient to consider the case which lower bounds $\gamma(\theta,\phi)'$, the dampening term of the general channel, which is the channel: $\cos^2\left(\frac{\phi}{2}\right)\mathcal{L}(\hat{Z})+\sin^2\left(\frac{\phi}{2}\right)\mathcal{L}(\hat{X})$, where $\mathcal{L}(\hat{Z})$ and $\mathcal{L}(\hat{X})$ are the dephasing and bit flip channels from before. This channel does not provide all possible noise terms that the general channel possesses; thus, it is less noisy than the general case.  

Following the same derivations from Theorems~\ref{Theorem1} and~\ref{Theorem2}, we see that our dampening function $\gamma(\theta,\phi), 0\leq \phi \leq \pi,$ for the approximate case is given as:
\begin{align}
    \label{general gamma}
    \gamma(\theta,\phi) &= (1- (1-p\theta)^{N}) - (1-(1-p\theta)^N) \times \nonumber \\
     & \ \ \left(\sin^2\left(\frac{\phi}{2}\right)-\cos^2\left(\frac{\phi}{2}\right)\right)-2\cos^2\left(\frac{\phi}{2}\right)\tilde{W}_{C^{\perp}}.
\end{align}
We see that we recover the dephasing case of $\gamma(\theta)$ for $\gamma(\theta,0)$, and the bit-flip case for $\gamma(\theta,\pi)$, respectively. We wish to make \eqref{general gamma} as close to zero as possible, which implies that
\begin{align}
\label{general robustness bound}
 \tilde{W}_{C^{\perp}} & \leq \frac{(1- (1-p\theta)^{N})}{2 \cos^{2}\frac{\phi}{2}} \left[ 1 - \sin^2\frac{\phi}{2} + \cos^2\frac{\phi}{2} \right] \nonumber \\
%
%
 & = 1-(1-p\theta)^{N},
 \end{align}

However, for \eqref{general robustness bound} to be an equality, we see from \eqref{robustness} that $W_{C^{\perp},k}=\binom{N}{k}$, which would mean the code $C$ is invalid by the quantum singleton bound. Thus, we see that only a trivial code with dual code consisting of all weight-1 vectors saturates the bound, which recovers our previous result. This means that the only way we can achieve the HL is to be subject to a pure bit-flip channel, or when $\phi=\pi$. Since the probe state's ability to suppress channel noise parallel to the Hamiltonian is bounded by the Robustness bound regardless of $\phi$, the channel noise cannot be suppressed enough to recover the HL unless the Hamiltonian is fully orthogonal to the channel operators.
\end{IEEEproof}
\begin{remark}
We note that the RHS of Eqn.~\eqref{general robustness bound} is a lower bound to the exact dampening term for a general channel comprised of the operator $\hat{U}(\phi)=\cos\left(\frac{\phi}{2}\right)\hat{Z}+\sin\left(\frac{\phi}{2}\right)\hat{X}$. In Appendix~\ref{sec:AppenC} we derive the exact inequality and tie it to  \eqref{general robustness bound}.
\end{remark}

\subsection{Numerical Results}

Continuing with our example of using a GHZ state with our protocol from section ~\ref{sec:protocol}, we simulated the probe state evolution for the Lindbladian cases in Theorems~\ref{Theorem1}-\ref{Theorem3} respectively, where for Theorem~\ref{Theorem3}, we choose $\phi=\frac{\pi}{2}$, or the Hadamard channel. These simulations done in QuTiP \cite{qutip5} with $\theta=10^{-3}$, $p=0.05$, were used to plot the Cramer-Rao bound \eqref{CRB}, $\delta\theta$, as a function of time for each case for the $N=7$ GHZ state. We use QuTiP's \texttt{mesolve} function to evolve the probe state via the Lindbladian evolutions of the different channels, then calculate the probability to obtain the +1 syndrome for all stabilizers. We insert this probability distribution into~\eqref{cfi}, which is equal to the QFI since measuring the stabilizers yields the maximum Classical Fisher Information from our probability distribution \cite{PhysRevA.97.042337}. Using this approach, we also compared these results to the logical $\ket{0}$ state of the Steane code.

\begin{figure}[t]

\centering
\scalebox{0.35}{
    \input{CramerRaovtimeGHZ.pgf}
}
\caption{Cramer-Rao bound as a function of time for the $N=7$ GHZ state. Here, we see that only in the X-Channel case does the GHZ achieve the HL~\eqref{bpcfi}, with $W_{C^{\perp},2}=\binom{N}{2}$ in~\eqref{QFIpurs}. In the Dephasing and Hadamard Channel cases, it is not even reaching the SQL~\eqref{bpcfi}, with $W_{C^{\perp},2}=0$ in~\eqref{QFIpurs}.}
\label{fig:ghz}
\end{figure}

\begin{figure}[t]

\centering
\scalebox{0.35}{
    \input{CramerRaovtimeSteane.pgf}
}
\caption{Cramer-Rao bound as a function of time for the $\ket{0}_{L}$ state of the Steane code. The Steane code, due to its low $\mathcal{Q}_{\rm pure}$ given by \eqref{QFIpurs}, can only achieve the SQL for the X-Channel Case, as predicted by Theorem 2.}
\label{fig:steane}
\end{figure}

In Fig.~\ref{fig:ghz}, we notice that the Cramer-Rao bound governed by $\mathcal{Q}_{\rm pure}$ of the probe state is only achievable when the channel is orthogonal to the Hamiltonian, as in the case of the $\hat{X}$-Channel. In this case, since the GHZ state's $\mathcal{Q}_{\rm pure}$ is maximal, it can reach the HL, unlike the Steane code in Fig.~\ref{fig:steane}, which does not contain any weight-2 dual-codewords, meaning it can only approach the SQL. 

 Since the Robustness of the probe state prevents Heisenberg Scaling, finite rounds of active error correction will not achieve that scaling either. Still, they may help achieve an uncertainty closer to saturating the Cramer-Rao bound for the SQL case~\cite{Arrad-prl14,Kapourniotis-pra19}. However, it should be noted that the Heisenberg Scaling is attainable if the channel has orthogonal support to the Hamiltonian on a few qubits: i.e., a set of $\hat{X}$ acting on $w$ qubits for a Hamiltonian of the form \eqref{hamiltonian}. We again see that for this case, the Heisenberg Scaling is attainable for a probe state built out of a code $C$ with maximal number of weight-2 dual codewords. However, a physical system that obeys the HNLS condition is difficult to see in practice.

\section{Outlook}
\label{sec:outlook}
In this paper, we introduced a novel method for using the weight enumerator of the dual of a linear code $C$, $W_{C^{\perp}}$, to calculate the QFI of a probe state constructed from $C$. In the general case, we showed that the weight enumerator of the dual code of $C$ also determines the robustness bound, which determines if, under a general dephasing channel, the probe state can achieve Heisenberg-limited metrology. We find that the robustness bound leads to the HNLS condition holding for linear-code robust metrology. Our method for determining this result demonstrates that coding theory techniques can be an elegant tool for finding fundamental physics results. Although our result limits the efficacy of quantum error correction protocols for metrology applications, we note that in previous work, the Heisenberg Scaling was shown to be attainable for erasure channels both in the case of finite-round error correction and robust sensing regimes~\cite{Ouyang-pra23,Ouyang-arxiv22}, both for linear codes and permutation-invariant codes. This means such scaling could be attainable by utilizing erasure qubits using a similar setup to what is described here~\cite{Niroula-prl24}. Another exciting direction for metrology is using non-Markovian dynamics and non-linear Kerr interactions as sensing resources, which have already shown promise in surpassing SQL~\cite{Yang-commphy24,Zhang-qip25}. Regardless of which approach is considered, there are many open questions about which approach will be the most practical. 
Finally, we note that in distributed quantum metrology, graph states are typically seen as a path to optimal metrology~\cite{Shettell-prl20}, with this intuition typically arising from considering trivial, or truncated noise models. However, since graph states are usually linear quantum codes over $\mathbb{F}_{2}^{n}$, our results show that under more realistic noise, graph states are likely bounded by the SQL.

 \section*{Acknowledgements}
 We thank Christos Gagatsos and Kanu Sinha for their insights into our paper. Their expertise in quantum metrology proved invaluable. The work of O.~N. and N.~R. was supported by the U.S. Army Research Office Grant W911NF-24-1-0080.

\IEEEtriggeratref{29}

\bibliographystyle{IEEEtran}
\bibliography{cites}

\onecolumn

\begin{appendices}
\section{Proof of Theorem 1} 
\label{sec:AppenA}

\begin{lemma} 
\label{lemma:1}

Given a general dephasing channel and a probe state constructed from all codewords of linear code $C$, to first-order in $dt$, the density operator is given by:
\begin{equation}
\begin{aligned}
&\rho(t+dt)=\frac{1}{|C|}\sum_{\boldsymbol{x},\boldsymbol{y} \in C} \ket{\boldsymbol{x}}\bra{\boldsymbol{y}} -\frac{i\theta dt}{|C|}\sum_{\boldsymbol{x}\neq\boldsymbol{y},\boldsymbol{i}} (-1)^{\left<\boldsymbol{x},\boldsymbol{i}\right>}\ket{\boldsymbol{x}}\bra{\boldsymbol{y}}+(-1)^{\left<\boldsymbol{y},\boldsymbol{i}\right>+1}\ket{\boldsymbol{x}}\bra{\boldsymbol{y}} +\\ &\frac{dt}{|C|}\Biggr(\sum_{\boldsymbol{x}, \boldsymbol{y} \in C} -(2-2(1-p\theta)^{N}) \ket{\boldsymbol{x}}\bra{\boldsymbol{y}} + \frac{2dt}{|C|}\sum_{k>0, \boldsymbol{x}, \boldsymbol{y} \in C}^{N} (1-p\theta)^{N-k}(p\theta)^{k} W_{C^{\perp},k}\ket{\boldsymbol{x}}\bra{\boldsymbol{y}}\Biggr)\\&+\frac{2dt}{|C|}\sum_{k>0, \boldsymbol{x}, \boldsymbol{y} \in C}^{N} (1-p\theta)^{N-k}(p\theta)^{k} W_{\boldsymbol{x}\oplus\boldsymbol{y}^{\perp},k} \ket{\boldsymbol{x}}\bra{\boldsymbol{y}}
\end{aligned}
\end{equation}
\end{lemma}

\begin{IEEEproof}[Proof of Lemma~\ref{lemma:1}]
 Let our probe state be given by $\rho=\frac{1}{|C|}\sum_{\boldsymbol{x}, \boldsymbol{y} \in C}\ket{\boldsymbol{x}}\bra{\boldsymbol{y}}$, where $C$ is a non-trivial, linear code. We subject our probe state to the following evolution:
 {\small
 \begin{equation}
     \label{eq:nqcase}
     \frac{\partial \rho}{\partial t}=-\frac{i}{\hbar}[\hat{H},\rho]\theta - \rho+\sum_{k=0, \boldsymbol{j}\in \mathcal{I}}^N (1-p(\theta))^{N-k}(p(\theta))^{k} E(\boldsymbol{0},\boldsymbol{j}) \rho E(\boldsymbol{0},\boldsymbol{j}) ,
 \end{equation}
 }
 where $\boldsymbol{j}\in \{0,1\}^N, |\boldsymbol{j}|=k$, and $\mathcal{I}$ 
 is the set of all length $N$ binary vectors with weight $k$, and the $-\rho$ term comes from the anti-commutator term from the Lindbladian.   We also use the following Hamiltonian:
 \begin{equation}
 \label{otherh}
 \hat{H}=\sum_{\boldsymbol{i}} \hat{Z}_{\boldsymbol{i}},  
\end{equation}
where $\boldsymbol{i}$ indexes over all weight 1, length $N$ binary vectors.
To first order, we find our density operator $\rho(t)$ as:
\begin{equation}
    \rho(t+dt)=\rho+ \frac{\partial \rho}{\partial t} dt
\end{equation}
 We now take the trace with respect to some projector $\Pi$ from our POVM: which we define to be the sum of all of our code's $\hat{X}$ stabilizers:
\begin{equation}
\label{eq:projector}
\Pi=\frac{1}{|C|}\sum_{\boldsymbol{s}\in C }E(\boldsymbol{s},\boldsymbol{0})
\end{equation}
We first evaluate the commutator given in the first term of \eqref{eq:nqcase}. This evaluates to:

\begin{equation}
\label{eq:ncommutator}
[H,\rho]=\frac{1}{|C|}\sum_{\boldsymbol{x},\boldsymbol{y}\in C, \boldsymbol{i}} (-1)^{\left<\boldsymbol{x},\boldsymbol{i}\right>}\ket{\boldsymbol{x}}\bra{\boldsymbol{y}}+(-1)^{\left<\boldsymbol{y},\boldsymbol{i}\right>+1}\ket{\boldsymbol{x}}\bra{\boldsymbol{y}},
\end{equation}
where we see that for this to be non-zero then $\left<\boldsymbol{x},\boldsymbol{i}\right>=\left<\boldsymbol{y},\boldsymbol{i}\right>+1$. 
For this to be true, $\boldsymbol{x}\oplus\boldsymbol{y} \neq \boldsymbol{0}$, so only the off-diagonal terms survive. We now evaluate the channel term of \eqref{eq:nqcase}. First, we note that $p(\theta)$ is small, so we approximate it as $p\theta$. Next, we evaluate the dephasing channel as:

\begin{align}
\label{eq:ndepterm}
&\sum_{\boldsymbol{x}, \boldsymbol{y} \in C, k=0, \boldsymbol{j}\in \mathcal{I}}^N (1-p\theta)^{N-k}(p\theta)^{k} E(\boldsymbol{0},\boldsymbol{j}) \rho E(\boldsymbol{0},\boldsymbol{j})= \\ \nonumber &\sum_{\boldsymbol{x}, \boldsymbol{y} \in C, k=0, \boldsymbol{j}\in \mathcal{I}}^N (1-p\theta)^{N-k}(p\theta)^{k} (-1)^{\left<\boldsymbol{x}\oplus\boldsymbol{y},\boldsymbol{j}\right>} \ket{\boldsymbol{x}}\bra{\boldsymbol{y}}.
\end{align}

 We expand this sum over $k$ giving: 
\begin{align}
 \label{expandederror}
 \nonumber &\sum_{\boldsymbol{x}, \boldsymbol{y} \in C, k=0, \boldsymbol{j}\in \mathcal{I}}^N (1-p\theta)^{N-k}(p\theta)^{k} (-1)^{\left<\boldsymbol{x}\oplus\boldsymbol{y},\boldsymbol{j}\right>} \ket{\boldsymbol{x}}\bra{\boldsymbol{y}} = \\ &\nonumber  \sum_{\boldsymbol{x}, \boldsymbol{y} \in C} \bigg[ \left( 1-p\theta+(p\theta)^2 + \cdots+ (-1)^{N}(p\theta)^N \right) + \\ &\nonumber \cdots\ + \left( (p\theta)^k-(p\theta)^{k+1}+\cdots+(-1)^{N-k}(p\theta)^{N} \right) \sum_{\boldsymbol{j}_{k}}(-1)^{\left<\boldsymbol{x}\oplus\boldsymbol{y},\boldsymbol{j}_{k}\right>}\\&+\cdots+(p\theta)^{N}\sum_{\boldsymbol{j}_{N}}(-1)^{\left<\boldsymbol{x}\oplus\boldsymbol{y},\boldsymbol{j}_{N}\right>} \bigg]  \ket{\boldsymbol{x}}\bra{\boldsymbol{y}}.
\end{align}

 Noticing a lack of a phase on the $k=0$ term, we rewrite \eqref{expandederror} as: 

\begin{align}
\label{expandederror2}
 \sum_{\boldsymbol{x}, \boldsymbol{y} \in C, k=0, \boldsymbol{j}\in \mathcal{I}}^N &(1-p\theta)^{N-k}(p\theta)^{k} (-1)^{\left<\boldsymbol{x}\oplus\boldsymbol{y},\boldsymbol{j}\right>} \ket{\boldsymbol{x}}\bra{\boldsymbol{y}} = \nonumber\\ \sum_{\boldsymbol{x}, \boldsymbol{y} \in C} &(1-p\theta)^{N} \ket{\boldsymbol{x}}\bra{\boldsymbol{y}} + \sum_{k>0}^{N} (1-p\theta)^{N-k}(p\theta)^{k}\sum_{\boldsymbol{j}_{k}}\sum_{\boldsymbol{x}, \boldsymbol{y} \in C} (-1)^{\left<\boldsymbol{x}\oplus\boldsymbol{y},\boldsymbol{j}_{k}\right>} \ket{\boldsymbol{x}}\bra{\boldsymbol{y}}
\end{align}

  Here we notice that either $\boldsymbol{j}_{k} \in C^{\perp}$, or $\boldsymbol{j}_{k} \notin C^{\perp}$.  We, therefore, write the second term of \eqref{expandederror2} as:
  
 \begin{align}
\label{enumeratorterm}
 \nonumber &\sum_{k>0}^{N} (1-p\theta)^{N-k}(p\theta)^{k}\sum_{\boldsymbol{j}_{k}}\sum_{\boldsymbol{x}, \boldsymbol{y} \in C} (-1)^{\left<\boldsymbol{x}\oplus\boldsymbol{y},\boldsymbol{j}_{k}\right>} \ket{\boldsymbol{x}}\bra{\boldsymbol{y}}= \\ & \sum_{k>0}^{N} (1-p\theta)^{N-k}(p\theta)^{k}\Biggr(\sum_{\boldsymbol{j}_{k}\in C^{\perp}}\sum_{\boldsymbol{x}, \boldsymbol{y} \in C} \ket{\boldsymbol{x}}\bra{\boldsymbol{y}} +\sum_{\boldsymbol{j}_{k}\notin C^{\perp}}\sum_{\boldsymbol{x}, \boldsymbol{y} \in C} (-1)^{\left<\boldsymbol{x}\oplus\boldsymbol{y},\boldsymbol{j}_{k}\right>} \ket{\boldsymbol{x}}\bra{\boldsymbol{y}}\Biggr)
\end{align}

  The sum $\sum_{\boldsymbol{j}_{k}\in C^{\perp}} (1)$ enumerates the number of weight $k$ dual codewords in the code, given by the MacWilliams codeword enumerator $W_{C^{\perp},k}$. But for the sum $\sum_{\boldsymbol{j}_{k} \notin C^{\perp}} (1)$, we need to be careful as not every term carries the same sign. We now focus on this term: first we swap the order of the sums, and treat each $\boldsymbol{x}\oplus\boldsymbol{y}$ as a one-dimensional code $C_{\boldsymbol{x}\oplus\boldsymbol{y}}$. If $\boldsymbol{j}_{k} \in C_{\boldsymbol{x}\oplus\boldsymbol{y}}^{\perp}$, then the sign is $1$, otherwise it is $-1$. We can now enumerate the number of weight $k$ dual codewords of $C_{\boldsymbol{x}\oplus\boldsymbol{y}}$ with $W_{\boldsymbol{x}\oplus\boldsymbol{y}^{\perp},k}$, which does not include $\boldsymbol{j}_{k} \in C^{\perp}$.  Thus we can rewrite \eqref{enumeratorterm} as:


\begin{align}
\label{enumeratorterm3}
\sum_{k>0}^{N} &(1-p\theta)^{N-k}(p\theta)^{k} \left[ \sum_{\boldsymbol{j}_{k}\in C^{\perp}}\sum_{\boldsymbol{x}, \boldsymbol{y} \in C} \ket{\boldsymbol{x}}\bra{\boldsymbol{y}} +\sum_{\boldsymbol{j}_{k}\notin C^{\perp}}\sum_{\boldsymbol{x}, \boldsymbol{y} \in C}(-1)^{\left<\boldsymbol{x}\oplus\boldsymbol{y},\boldsymbol{j}_{k}\right>} \ket{\boldsymbol{x}}\bra{\boldsymbol{y}} \right]= \nonumber \\
\sum_{k>0, \boldsymbol{x}, \boldsymbol{y} \in C}^{N} &(1-p\theta)^{N-k}(p\theta)^{k} W_{C^{\perp},k}\ket{\boldsymbol{x}}\bra{\boldsymbol{y}} +\sum_{k>0, \boldsymbol{x}, \boldsymbol{y} \in C}^{N} (1-p\theta)^{N-k}(p\theta)^{k} W_{\boldsymbol{x}\oplus \boldsymbol{y}^{\perp},k}\ket{\boldsymbol{x}}\bra{\boldsymbol{y}}\nonumber \\ 
\qquad - \sum_{k>0, \boldsymbol{x}, \boldsymbol{y} \in C}^{N} &(1-p\theta)^{N-k}(p\theta)^{k} \left( \binom{N}{k} - W_{C^{\perp},k}- W_{\boldsymbol{x}\oplus\boldsymbol{y}^{\perp},k} \right) \ket{\boldsymbol{x}}\bra{\boldsymbol{y}}=\nonumber \\
-\sum_{k>0, \boldsymbol{x}, \boldsymbol{y} \in C}^{N} &(1-p\theta)^{N-k}(p\theta)^{k} \binom{N}{k} \ket{\boldsymbol{x}}\bra{\boldsymbol{y}} \nonumber \\
\qquad + 2 \sum_{k>0, \boldsymbol{x}, \boldsymbol{y} \in C}^{N} &(1-p\theta)^{N-k}(p\theta)^{k} W_{C^{\perp},k}\ket{\boldsymbol{x}}\bra{\boldsymbol{y}} \nonumber \\ 
\qquad + 2 \sum_{k>0, \boldsymbol{x}, \boldsymbol{y} \in C}^{N} &(1-p\theta)^{N-k}(p\theta)^{k} W_{\boldsymbol{x}\oplus\boldsymbol{y}^{\perp},k} \ket{\boldsymbol{x}}\bra{\boldsymbol{y}}
\end{align}

Combining with the first term of \eqref{expandederror2}, we get:

\begin{align}   
\label{eq:evaldep}
 \sum_{\boldsymbol{x}, \boldsymbol{y} \in C, k=0, \boldsymbol{j}\in \mathcal{I}}^N &(1-p\theta)^{N-k}(p\theta)^{k}  E(\boldsymbol{0},\boldsymbol{j}) \rho E(\boldsymbol{0},\boldsymbol{j})= \nonumber \\ \nonumber \sum_{\boldsymbol{x}, \boldsymbol{y} \in C} -&(1-2(1-p\theta)^{N}) \ket{\boldsymbol{x}}\bra{\boldsymbol{y}} + 2\sum_{k>0, \boldsymbol{x}, \boldsymbol{y} \in C}^{N} (1-p\theta)^{N-k}(p\theta)^{k} W_{C^{\perp},k}\ket{\boldsymbol{x}}\bra{\boldsymbol{y}}\nonumber\\+2\sum_{k>0, \boldsymbol{x}, \boldsymbol{y} \in C}^{N} &(1-p\theta)^{N-k}(p\theta)^{k} W_{\boldsymbol{x}\oplus\boldsymbol{y}^{\perp},k}\ket{\boldsymbol{x}}\bra{\boldsymbol{y}}
\end{align}

Thus, our density operator now has the form:
\begin{equation}
\begin{aligned}   
\label{eq:fullop}
&\rho(t+dt)=\frac{1}{|C|}\sum_{\boldsymbol{x},\boldsymbol{y} \in C} \ket{\boldsymbol{x}}\bra{\boldsymbol{y}} -\frac{i\theta dt}{|C|}\sum_{\boldsymbol{x}\neq\boldsymbol{y},\boldsymbol{i}} (-1)^{\left<\boldsymbol{x},\boldsymbol{i}\right>}\ket{\boldsymbol{x}}\bra{\boldsymbol{y}}+(-1)^{\left<\boldsymbol{y},\boldsymbol{i}\right>+1}\ket{\boldsymbol{x}}\bra{\boldsymbol{y}} +\\ &\frac{dt}{|C|}\Biggr(\sum_{\boldsymbol{x}, \boldsymbol{y} \in C} -(2-2(1-p\theta)^{N}) \ket{\boldsymbol{x}}\bra{\boldsymbol{y}} + \frac{2dt}{|C|}\sum_{k>0, \boldsymbol{x}, \boldsymbol{y} \in C}^{N} (1-p\theta)^{N-k}(p\theta)^{k} W_{C^{\perp},k}\ket{\boldsymbol{x}}\bra{\boldsymbol{y}}\Biggr)\\&+\frac{2dt}{|C|}\sum_{k>0, \boldsymbol{x}, \boldsymbol{y} \in C}^{N} (1-p\theta)^{N-k}(p\theta)^{k} W_{\boldsymbol{x}\oplus\boldsymbol{y}^{\perp},k} \ket{\boldsymbol{x}}\bra{\boldsymbol{y}}
\end{aligned}
\end{equation}

We label the sum over $k$ of our second to last term as: $\tilde{W}_{C^{\perp}}$, which we define as our Robustness.
\end{IEEEproof} 
\begin{lemma}
\label{ref:lemma2}
The probability $p\left(\Pi=1|\rho(t+dt)\right)$ is  approximately given by:
\begin{equation}
p\left(\Pi=1|\rho(t+dt)\right)= \mathrm{Tr}(\rho(t+dt)\Pi)\approx Ae^{-\gamma(\theta)(t+dt)}\cos(\sqrt{\mathcal{Q}_{\rm pure}}\theta (t+dt)) +B,
\end{equation}
which holds for general codes up to second order, but is exact for the GHZ state.
\end{lemma}
\begin{IEEEproof}[Proof of Lemma~\ref{ref:lemma2}]
We now evaluate the product of $\rho(t+dt)$ with our projector $\Pi$, given by \eqref{eq:projector}.

\begin{align}
\label{eq:evalprojector}
\rho(t+dt)\Pi-\rho\Pi=-\frac{i\theta dt}{|C^2|}&\sum_{\boldsymbol{x}\neq\boldsymbol{y},\boldsymbol{s}\in C,\boldsymbol{i}} (-1)^{\left<\boldsymbol{x},\boldsymbol{i}\right>}\ket{\boldsymbol{x}}\bra{\boldsymbol{y}} E(\boldsymbol{s},\boldsymbol{0})+(-1)^{\left<\boldsymbol{y},\boldsymbol{i}\right>+1}\ket{\boldsymbol{x}}\bra{\boldsymbol{y}} E(\boldsymbol{s},\boldsymbol{0})\nonumber \\
-\frac{dt}{|C^2|} \Biggr( \sum_{\boldsymbol{x},\boldsymbol{y}, \boldsymbol{s} \in C}&(2-2(1-p\theta)^{N}) \ket{\boldsymbol{x}}\bra{\boldsymbol{y}} E(\boldsymbol{s},\boldsymbol{0})- \frac{dt}{|C^2|}\sum_{\boldsymbol{x}, \boldsymbol{y}, \boldsymbol{s} \in C} \tilde{W}_{C^{\perp}} \ket{\boldsymbol{x}}\bra{\boldsymbol{y}}E(\boldsymbol{s},\boldsymbol{0}) \Biggr) + \nonumber\\ 
\frac{dt}{|C^2|}\sum_{k>0, \boldsymbol{x}, \boldsymbol{y}, \boldsymbol{s} \in C}^{N} &(1-p\theta)^{N-k}(p\theta)^{k} W_{\boldsymbol{x}\oplus\boldsymbol{y}^{\perp},k}\ket{\boldsymbol{x}}\bra{\boldsymbol{y}}E(\boldsymbol{s},\boldsymbol{0}) ,
\end{align}
where $\ket{\boldsymbol{x}}\bra{\boldsymbol{y}}E(\boldsymbol{s},\boldsymbol{0})=\ket{\boldsymbol{x}}\bra{\boldsymbol{y}\oplus\boldsymbol{s}}$. Taking the trace of \eqref{eq:evalprojector} yields:

\begin{align}
\label{aaaa}
&\mathrm{Tr}(\rho(t+dt)\Pi)-tr(\rho\Pi)=-\frac{i\theta dt}{|C^2|}\sum_{\boldsymbol{x}\neq\boldsymbol{y},\boldsymbol{s} \in C,\boldsymbol{i}} (-1)^{\left<\boldsymbol{x},\boldsymbol{i}\right>}\delta_{\boldsymbol{x},\boldsymbol{y}\oplus\boldsymbol{s}}+(-1)^{\left<\boldsymbol{y},\boldsymbol{i}\right>+1}\delta_{\boldsymbol{x},\boldsymbol{y}\oplus\boldsymbol{s}}\nonumber \\&-\frac{dt}{|C^2|}\sum_{\boldsymbol{x}, \boldsymbol{y}, \boldsymbol{s} \in C} ((2-2(1-p\theta)^{N}) -2\tilde{W}_{C^{\perp}})\delta_{\boldsymbol{x},\boldsymbol{y}\oplus\boldsymbol{s}} \nonumber \\&+\frac{dt}{|C^2|}\sum_{k>0, \boldsymbol{x}, \boldsymbol{y}, \boldsymbol{s} \in C}^{N} (1-p\theta)^{N-k}(p\theta)^{k} W_{\boldsymbol{x}\oplus\boldsymbol{y}^{\perp},k}\delta_{\boldsymbol{x},\boldsymbol{y}\oplus\boldsymbol{s}} ,
\end{align}

We can now evaluate the Kronecker Deltas and write \eqref{aaaa} in terms of a sum over all possible $X$ stabilizers, as we evaluate the sum over $\boldsymbol{s}$. This allows us to rewrite the sum in terms of $\boldsymbol{y}$ and $\boldsymbol{s}$. We now rewrite the first term to show how it vanishes. 
\begin{equation}
\label{eq:firsttermvanish1}
\begin{aligned}
\sum_{\boldsymbol{x}\neq\boldsymbol{y},\boldsymbol{s} \in C,\boldsymbol{i}} (-1)^{\left<\boldsymbol{x},\boldsymbol{i}\right>}\delta_{\boldsymbol{x},\boldsymbol{y}\oplus\boldsymbol{s}}=\sum_{\boldsymbol{i}}\sum_{\boldsymbol{y} \in C,}(-1)^{\left<\boldsymbol{y},\boldsymbol{i}\right>}\Biggr(\sum_{\boldsymbol{s} \in C\setminus\{\boldsymbol{0}\}}(-1)^{\left<\boldsymbol{s},\boldsymbol{i}\right>}-1\Biggr)
\end{aligned}
\end{equation}
Given the Linearity of $C$, any two codewords $\boldsymbol{x},\boldsymbol{y} \in C$, $\boldsymbol{x}\oplus\boldsymbol{y} \in C$. Since each $\boldsymbol{i}$ has weight one at i, we look at the index i. There exists some $\boldsymbol{c} \in C$ such that $c_{i}=1$, such that for each $\boldsymbol{x}\in C$, such that $x_{i}=0$, $\boldsymbol{x}\oplus\boldsymbol{c} \in C$ has $(\boldsymbol{x}\oplus\boldsymbol{c})_{i}=1$. Exactly $\frac{|C|}{2}$ codewords have $c_{i}=0$, thus the inner sum evaluates to $-|C|$. By the same logic, the outer sum in $\boldsymbol{y}$ must be zero since exactly half of the codewords in $C$ have $c_{i}=1$, while the other half has $c_{i}=0$. Thus, the term vanishes. To first order, we then see that our expansion evaluates to:

\label{eq:tracenq}
\begin{align}
&\nonumber \mathrm{Tr}(\rho(t+dt)\Pi)-\mathrm{Tr}(\rho\Pi)=\\&-\frac{dt}{|C^2|}\sum_{\boldsymbol{y},\boldsymbol{s} \in C}( (2-2(1-p\theta)^{N})-2\tilde{W}_{C^{\perp}}) +\frac{dt}{|C^2|}\sum_{k>0}^{N}\sum_{\boldsymbol{y}, \boldsymbol{s} \in C}(1-p\theta)^{N-k}(p\theta)^{k} W_{\boldsymbol{s}^{\perp},k},
\end{align}

Since our expansion to first order includes no signal term, we use a second order expansion. We now define:
\begin{equation}
     \label{gamma}\
     \gamma(\theta)=(2-2(1-p\theta)^N)-2\tilde{W}_{C^{\perp}}-2\sum_{k>0}^{N}\sum_{\boldsymbol{s} \in C}(1-p\theta)^{N-k}(p\theta)^{k} W_{\boldsymbol{s}^{\perp},k}
 \end{equation}
 and our second order expansion is approximately:
 \begin{equation}
\label{secondorder}
\begin{aligned}
\mathrm{Tr}(\rho(t+dt)\Pi) \approx 1-\gamma(\theta)dt+\left(-\theta (\mathcal{Q}_{\rm pure})+\frac{\gamma(\theta)^2}{2}\right)dt^2+\mathcal{O}(dt^3).
\end{aligned}
\end{equation}
We recognize this as a dampened cosine function. This approximation holds to second order for general codes, but for the GHZ state a dampened cosine function is the exact solution. Since as previously shown has higher QFI than any other probe state given $N$ qubits, it is sufficient to consider this case. Thus, our final form for our probability distribution reads:

\begin{equation}
\label{probabilitymeasure}
\begin{aligned}
p\left(\Pi=1|\rho(t+dt)\right) = \mathrm{Tr}(\rho(t+dt)\Pi)=Ae^{-\gamma(\theta)(t+dt)}\cos(\sqrt{\mathcal{Q}_{\rm pure}}\theta (t+dt)) +B,
\end{aligned}
\end{equation}
 where we choose  $A,B=\frac{1}{2}$ for our purposes here. However, we must reexamine $\gamma(\theta)$. The last term in \eqref{gamma} can cause $\gamma(\theta)<0$, which would mean that our noise channel amplifies our signal instead of dampening it. Such behavior is non-physical, so we must rescale \eqref{gamma} as:
 \begin{equation}
 \gamma(\theta) \rightarrow   \gamma(\theta)+2\sum_{k>0}^{N}\sum_{\boldsymbol{s} \in C^{\perp}}(1-p\theta)^{N-k}(p\theta)^{k} W_{\boldsymbol{s}^{\perp},k} = (2-2(1-p\theta)^N)-2\tilde{W}_{C^{\perp}}
 \end{equation}
 This shows us that our scaling behavior depends on the derived quantity $\tilde{W}_{C^{\perp}}$. 
\end{IEEEproof}
\begin{IEEEproof}[Proof of Theorem 1]
From our probability distribution \eqref{probabilitymeasure}, we can write the variance of our measurement as: 
\begin{equation}
    \delta \tilde{\theta} \geq \frac{1}{\sqrt{F(\rho,\Pi)}},
\end{equation}
Where our Classical Fisher Information, CFI,  is given by:
\begin{equation}
\label{cfi}
   F(\rho,\Pi)=\frac{1}{p\left(\Pi=1|\rho(t+dt)\right)}\left(\frac{\partial p\left(\Pi=1|\rho(t+dt)\right)}{\partial \theta}\right)^2+\frac{1}{(1-p\left(\Pi=1|\rho(t+dt)\right)}\left(\frac{\partial (1-p\left(\Pi=1|\rho(t+dt)\right))}{\partial \theta}\right)^2,
\end{equation}
Inserting \eqref{probabilitymeasure} into \eqref{cfi}, we see that:
\begin{equation}
    \label{cfiinvert}
    \frac{1}{F(\rho,\Pi)}=\frac{e^{2\gamma(\theta)(t+dt)}-\cos^{2}(\sqrt{\mathcal{Q}_{\rm pure}}\theta (t+dt))}{(dt+t)^2\left(\sqrt{\mathcal{Q}_{\rm pure}}\sin(\sqrt{\mathcal{Q}_{\rm pure}}\theta (t+dt))+\frac{\partial \gamma(\theta)}{\partial \theta}\cos(\sqrt{\mathcal{Q}_{\rm pure}}\theta (t+dt))\right)^2}=\frac{1}{\mathcal{Q}(\rho,\Pi)}
\end{equation}

For us to approximately recover the  Heisenberg Limit, we must let $\frac{\partial \gamma(\theta)}{\partial \theta}=0$. This implies that:
\begin{equation}
    \label{new limit}
    \tilde{W}_{C^{\perp}}\leq 1-(1-p\theta)^{N},
\end{equation}
where the bound is saturated by $W_{k}^{\perp}(z)=\binom{N}{k}, \forall k>0$. But this violates the quantum singleton bound, which means saturating this bound is unachievable. We call this quantity the robustness of the code.
Now, we can choose a code that approaches this bound to minimize the variance of the measurement. We see from the form of $\tilde{W}_{C^{\perp}}$ that maximizing the number of weight $1$ dual codewords will achieve the HL.  However, since the codewords of $C$ must commute with all of its weight one dual codewords, each codeword must not have support on each weight one dual codeword for this to be true. Since the HL is reached by having $N$ weight one dual codewords, the only allowed state is the all zeros state, which implies that our code $C$ must be trivial and cannot accumulate any signal. This is a contradiction. Since the GHZ state has higher QFI than any other probe state from a linear code $C$, it is not possible for any nontrivial linear code $C$ to achieve Heisenberg Scaling for this channel. It is also clear that any number of rounds of active error correction will further reduce the signal collected by any probe state in this scenario, as the action of \eqref{otherh} on $\rho$  is indistinguishable from a $k=1$ phase error from the channel. This means that this result holds for any number of rounds of active error correction.  
 \end{IEEEproof}
 \section{Proof of Theorem 2}
 \label{sec:AppenB}
 \begin{IEEEproof}[Proof of Theorem 2]
 We begin with the following channel:
\begin{equation}
     \label{eq:nqbitcase}
     \frac{\partial \rho}{\partial t}=-\frac{i}{\hbar}[\hat{H},\rho]\theta - \rho+\sum_{k=0, \boldsymbol{j}\in \mathcal{I}}^N (1-p(\theta))^{N-k}(p(\theta))^{k} E(\boldsymbol{j},\boldsymbol{0}) \rho E(\boldsymbol{j},\boldsymbol{0}) ,
 \end{equation}
where again the Hamiltonian is given by \eqref{otherh}, our POVM is given by \eqref{eq:projector}, and our probe state is given by the form in Appendix~\ref{sec:AppenA}. Since the derivation for the signal term is identical to the dephasing case, we begin from the noise term:
 \begin{align}
 \label{bitflipnoise}
 \nonumber \sum_{\boldsymbol{x}, \boldsymbol{y} \in C, k=0, \boldsymbol{j}\in \mathcal{I}}^N &(1-p\theta)^{N-k}(p\theta)^{k} E(\boldsymbol{j},\boldsymbol{0})\ket{\boldsymbol{x}}\bra{\boldsymbol{y}}E(\boldsymbol{j},\boldsymbol{0})=\sum_{\boldsymbol{x}, \boldsymbol{y} \in C} (1-p\theta)^{N} \ket{\boldsymbol{x}}\bra{\boldsymbol{y}} \\+\sum_{\boldsymbol{x}, \boldsymbol{y} \in C, k>0, \boldsymbol{j}_{k}\in \mathcal{I}}^N &(1-p\theta)^{N-k}(p\theta)^{k}\ket{\boldsymbol{x}\oplus \boldsymbol{j}_{k}}\bra{\boldsymbol{y}\oplus \boldsymbol{j}_{k}}
\end{align}
Again, we notice that either $\boldsymbol{j}_{k} \in C^{\perp}$, or $\boldsymbol{j}_{k} \notin C^{\perp}$. Since $\boldsymbol{x}, \boldsymbol{y}$ span all the codewords of $C$, and for any dual codeword $\boldsymbol{z}$ and code word $\boldsymbol{x}$, $ \boldsymbol{z} \oplus \boldsymbol{x} \in C$, we can use the dual code weight enumerator again to enumerate the number of $\boldsymbol{j}_{k} \in C^{\perp}$.
\begin{equation}
\begin{aligned}
\label{bitflipenum}
    &\sum_{\boldsymbol{x}, \boldsymbol{y} \in C, k>0, \boldsymbol{j}_{k}\in \mathcal{I}}^N (1-p\theta)^{N-k}(p\theta)^{k}\ket{\boldsymbol{x}\oplus \boldsymbol{j}_{k}}\bra{\boldsymbol{y}\oplus \boldsymbol{j}_{k}} =\\&\sum_{\boldsymbol{x}, \boldsymbol{y} \in C, k>0}^N (1-p\theta)^{N-k}(p\theta)^{k}\Biggr(\sum_{\boldsymbol{j}_{k}\in C^{\perp}} \ket{\boldsymbol{x}\oplus \boldsymbol{j}_{k}}\bra{\boldsymbol{y}\oplus \boldsymbol{j}_{k}}+\sum_{\boldsymbol{j}_{k}\notin C^{\perp}}\ket{\boldsymbol{x}\oplus \boldsymbol{j}_{k}}\bra{\boldsymbol{y}\oplus \boldsymbol{j}_{k}} \Biggr) 
\end{aligned}
\end{equation}
Taking the trace of the product of the first term of the RHS with our POVM, we see that:
\begin{equation}
\label{incodetracebit}
    \begin{aligned}
    &\mathrm{Tr}\left(\sum_{\boldsymbol{x}, \boldsymbol{y} \in C, k>0}^N (1-p\theta)^{N-k}(p\theta)^{k} \Biggr(\sum_{\boldsymbol{j}_{k}\in C^{\perp}}\ket{\boldsymbol{x}\oplus \boldsymbol{j}_{k}}\bra{\boldsymbol{y}\oplus \boldsymbol{j}_{k}} \Biggr) \Pi\right)= \\&\frac{1}{|C^2|}\sum_{k>0, \boldsymbol{x}, \boldsymbol{y} \in C, s \in C}^{N} (1-p\theta)^{N-k}(p\theta)^{k} W_{C^{\perp},k} \delta_{\boldsymbol{x},\boldsymbol{y}\oplus \boldsymbol{s}}= \\
    &\sum_{k>0}^{N} (1-p\theta)^{N-k}(p\theta)^{k} W_{C^{\perp},k} 
    \end{aligned}
\end{equation}
For the $\boldsymbol{j}_{k} \notin C^{\perp}$, we consider the set of one dimensional codes $C_{\boldsymbol{x}}, C_{\boldsymbol{y}}$. There are two cases: either $\boldsymbol{j}_{k} \in C_{\boldsymbol{x}}^{\perp} \cap C_{\boldsymbol{y}}^{\perp}$ or not. We count the number of $\boldsymbol{j}_{k}$ that are dual codewords of both as $W_{\boldsymbol{x}^{\perp},\boldsymbol{y}^{\perp},k}$. Since both $C_{\boldsymbol{x}}$, and $C_{\boldsymbol{y}}$ are one dimensional, we look at the action of $\boldsymbol{x} \oplus \boldsymbol{j}_{k}$. This sends $\boldsymbol{x}$ to an equivalent $\boldsymbol{x}'\in C_{\boldsymbol{x}}$ for each $\boldsymbol{j}_{k}$. Since the sum over $\boldsymbol{x}$ goes over all codewords, we consider a new code $C'$, made up of all codewords of each $C_{\boldsymbol{x}}$. Then the second term of \eqref{bitflipenum} can be written as:

\begin{align}
\label{notbitflipcodeword}
\sum_{\boldsymbol{x}, \boldsymbol{y} \in C, k>0}^N &(1-p\theta)^{N-k}(p\theta)^{k} \Biggr(\sum_{\boldsymbol{j}_{k}\notin C^{\perp}}\ket{\boldsymbol{x}\oplus \boldsymbol{j}_{k}}\bra{\boldsymbol{y}\oplus \boldsymbol{j}_{k}} \Biggr) = \nonumber \\\sum_{k>0, \boldsymbol{x}', \boldsymbol{y}' \in C'}^{N} &(1-p\theta)^{N-k}(p\theta)^{k} \ket{\boldsymbol{x}'}\bra{\boldsymbol{y}'},
\end{align}
where $\boldsymbol{x}', \boldsymbol{y}' \in C'$. Because each Pauli operator made of all Pauli $\hat{X}$ operators commutes with all other products of Pauli $\hat{X}$ operators, $W_{\boldsymbol{x}^{\perp},\boldsymbol{y}^{\perp},k}=\left(\binom{N}{k}- W_{C^{\perp},k}\right)$. Taking the trace with respect to \eqref{eq:projector}, the trace of~\eqref{notbitflipcodeword} can be written as: 

\label{traceofbitflip}
\begin{align}
    &\mathrm{Tr}\left(\sum_{\boldsymbol{x}, \boldsymbol{y} \in C, k>0}^N (1-p\theta)^{N-k}(p\theta)^{k} \Biggr(\sum_{\boldsymbol{j}_{k}\notin C^{\perp}}\ket{\boldsymbol{x}\oplus \boldsymbol{j}_{k}}\bra{\boldsymbol{y}\oplus \boldsymbol{j}_{k}} \Biggr) \Pi\right)=\nonumber \\&\frac{1}{|C^2|}\sum_{k>0, \boldsymbol{x}', \boldsymbol{y}' \in C', s \in C}^{N} (1-p\theta)^{N-k}(p\theta)^{k}  \delta_{\boldsymbol{x}',\boldsymbol{y}'\oplus \boldsymbol{s}} = \nonumber \\ &\frac{1}{|C^2|}\sum_{k>0, \boldsymbol{x}, \boldsymbol{y} \in C, s \in C}^{N} (1-p\theta)^{N-k}(p\theta)^{k} W_{\boldsymbol{x}^{\perp},\boldsymbol{y}^{\perp},k} \delta_{\boldsymbol{x},\boldsymbol{y}\oplus \boldsymbol{s}} =\nonumber  \\ &\frac{1}{|C^2|}\sum_{k>0, \boldsymbol{x}, \boldsymbol{y} \in C, s \in C}^{N} (1-p\theta)^{N-k}(p\theta)^{k} \left(\binom{N}{k}- W_{C^{\perp},k}\right) \delta_{\boldsymbol{x},\boldsymbol{y}\oplus \boldsymbol{s}}= \nonumber \\ &\sum_{k>0}^{N} (1-p\theta)^{N-k}(p\theta)^{k} \left(\binom{N}{k}- W_{C^{\perp},k}\right) 
\end{align}
Therefore, the trace \eqref{bitflipenum} can be written as: 
\begin{equation}
\begin{aligned}
 \mathrm{Tr}\left(\sum_{\boldsymbol{x}, \boldsymbol{y} \in C, k>0, \boldsymbol{j}_{k}\in \mathcal{I}}^N (1-p\theta)^{N-k}(p\theta)^{k} \ket{\boldsymbol{x}\oplus \boldsymbol{j}_{k}}\bra{\boldsymbol{y}\oplus \boldsymbol{j}_{k}}\Pi\right) &= \sum_{k>0}^N  (1-p\theta)^{N-k}(p\theta)^{k} \binom{N}{k}\\ &= 1-(1-p\theta)^{N},
 \end{aligned}
\end{equation}
Which means that the dampening due to noise here vanishes. The rest of the derivation follows the same logic as before, which leads to:
\begin{equation}
\label{probabilitymeasurebp}
\begin{aligned}
p\left(\Pi=1|\rho(t+dt)\right)=tr(\rho(t+dt)\Pi) \approx A\cos(\sqrt{\mathcal{Q}_{\rm pure}}\theta (t+dt)) +B,
\end{aligned}
\end{equation}
where $A=B=\frac{1}{2}$, and again holds to second-order for general codes, but is exact for the GHZ state for short time scales. Thus by inserting  \eqref{probabilitymeasurebp}
into \eqref{cfi}, we see that: 
\begin{equation}
    \label{bpcfi}
    \delta\theta \approx \sqrt{\frac{1}{(dt+t)^2\mathcal{Q}_{\rm pure}}},
\end{equation}
thus, we asymptotically achieve the HL as long as we maximize the number weight 2 dual codewords of our Code. 
 \end{IEEEproof}

\section{Deriving General Form of $\gamma(\theta,\phi)$}
\label{sec:AppenC}
We begin with the following channel:
\begin{equation}
     \label{eq:gencase}
     \frac{\partial \rho}{\partial t}=-\frac{i}{\hbar}[H,\rho]\theta - \rho+\sum_{k=0, \boldsymbol{j}\in \mathcal{I}}^N (1-p(\theta))^{N-k}(p(\theta))^{k} \hat{U}(\phi)^{\otimes \boldsymbol{j}} \rho \hat{U}^{\dagger}(\phi)^{\otimes \boldsymbol{j}},
 \end{equation}
 where $\hat{U}(\phi)=cos\left(\frac{\phi}{2}\right)\hat{Z}+\sin\left(\frac{\phi}{2}\right)\hat{X}$, the Hamiltonian is given by \eqref{otherh}, and the probe state is the same form as in Appendix~\ref{sec:AppenA}.
 We see that for \eqref{eq:gencase}, that for $\phi=0,\pi$, the channel reduces to the previous cases. Let us now focus on the noise term of the case where $0< \phi < \pi$. Here, we can expand the second term of \eqref{eq:gencase} as:
\begin{equation}
    \label{gennoiseexpand}
    \begin{aligned}
    &\sum_{k=0, \boldsymbol{j}\in \mathcal{I}}^N (1-p\theta)^{N-k}(p\theta)^{k} \hat{U}(\phi)^{\otimes \boldsymbol{j}} \rho \hat{U}^{\dagger}(\phi)^{\otimes \boldsymbol{j}}= (1-p\theta)^N+\sum_{k>0, \boldsymbol{j}\in \mathcal{I}}^N (1-p\theta)^{N-k}(p\theta)^{k} \epsilon(\rho)',\\
   & \epsilon(\rho)'= \sum_{\boldsymbol{l},\boldsymbol{m}\in \mathcal{J}} \cos\left(\frac{\phi}{2}\right)^{k-|\boldsymbol{l}|}\sin\left(\frac{\phi}{2}\right)^{|\boldsymbol{l}|}\cos\left(\frac{\phi}{2}\right)^{k-|\boldsymbol{m}|}\sin\left(\frac{\phi}{2}\right)^{|\boldsymbol{m}|}E(\boldsymbol{j}-\boldsymbol{l},\boldsymbol{l})\rho E(\boldsymbol{j}-\boldsymbol{m},\boldsymbol{m}),
    \end{aligned}
\end{equation}
where $\mathcal{J} $ is the set of all length $N$ binary vectors with weight up to k. Applying the same approach as in Appendix~\ref{sec:AppenA} for the $\hat{Z}$ case and Appendix~\ref{sec:AppenB} for the $\hat{X}$ case, taking the trace of the noise term yields a modified version of $\gamma(\theta,\phi)$:

\begin{align}
 \label{gammaprime}
&\gamma(\theta,\phi)'=1-(1-p\theta)^{N}+\sum_{k>0}^N  (1-p\theta)^{N-k}(p\theta)^{k} \sum_{l=0}^k \cos\left(\frac{\phi}{2}\right)^{2(k-l)} \sin\left(\frac{\phi}{2}\right)^{2l} \binom{N}{l}\binom{N}{k-l}-2\tilde{W}'_{C^{\perp}}, \nonumber \\
&\tilde{W}'_{C^{\perp}}=\sum_{k>0}^N  (1-p\theta)^{N-k}(p\theta)^{k} \sum_{l=0}^k \cos\left(\frac{\phi}{2}\right)^{2(k-l)}\sin\left(\frac{\phi}{2}\right)^{2l}\binom{N}{l}W_{C^{\perp},k-l} , \  0<\phi<\pi,
\end{align}

where we notice that $\tilde{W}_{C^{\perp}}'$ is equal to the second term when $W_{C^{\perp},k-l}=\binom{N}{k-l}$. We wish to show a relationship between the general case $\gamma(\theta,\phi)'$ and our approximate case in Theorem 3. To do this, we apply a re-scaling to  \eqref{gammaprime} by:
\begin{equation}
    \label{RG step}
    \gamma(\theta,\phi)' \rightarrow \gamma(\theta,\phi)' +\sum_{k>0}^N  (1-p\theta)^{N-k}(p\theta)^{k}  \left(-W_{C^{\perp},k}+\sum_{l=0}^k \cos\left(\frac{\phi}{2}\right)^{2(k-l)}\sin\left(\frac{\phi}{2}\right)^{2l}\binom{N}{l}W_{C^{\perp},k-l}\right),
\end{equation}
and since $\tilde{W}_{C^{\perp}}'$ is equal to the second term of \eqref{gammaprime} when $W_{C^{\perp},k-l}=\binom{N}{k-l}$, this implies that under our rescaling:

\begin{align}
&\gamma(\theta,\phi)' \approx \gamma(\theta,\phi)+\epsilon,\nonumber \\
&\lim_{W_{C^{\perp},k-l} \rightarrow \binom{N}{k-l}} \epsilon =0.
\end{align}
Thus, we see that $\tilde{W}_{C^{\perp}}\leq 1-(1-p\theta)^N+\epsilon$, $0<\phi<\pi$, which agrees with the conclusions of Theorem 3.
\end{appendices}


\end{document}